\documentclass[12pt,a4paper]{article}
\usepackage{amsmath}
\usepackage{latexsym}
\usepackage{amssymb}
\usepackage{graphicx,color}
\makeatletter
\def\rddots{\mathinner{\mkern1mu\raise\p@%
    \vbox{\kern7\p@\hbox{.}}\mkern2mu%
    \raise4\p@\hbox{.}\mkern2mu\raise7\p@\hbox{.}\mkern1mu}}
\makeatother
\makeatletter
\def\eqnarray{%
\stepcounter{equation}%
\let\@currentlabel=\theequation
\global\@eqnswtrue
\global\@eqcnt\z@
\tabskip\@centering
\let\\=\@eqncr
$$\halign to \displaywidth\bgroup\@eqnsel\hskip\@centering
$\displaystyle\tabskip\z@{##}$&\global\@eqcnt\@ne
\hfil$\displaystyle{{}##{}}$\hfil
&\global\@eqcnt\tw@$\displaystyle\tabskip\z@{##}$\hfil
\tabskip\@centering&\llap{##}\tabskip\z@\cr}
\makeatother
\begin{document}

\title{\sl Flow Representation of the Bose--Hubbard Hamiltonian : 
General Case}
\author{
  Kazuyuki FUJII
  \thanks{E-mail address : fujii@yokohama-cu.ac.jp }\quad and \ 
  Tatsuo SUZUKI
  \thanks{E-mail address : suzukita@aoni.waseda.jp ; 
  i027110@sic.shibaura-it.ac.jp }\\
  ${}^{*}$Department of Mathematical Sciences\\
  Yokohama City University\\
  Yokohama, 236--0027\\
  Japan\\
  ${}^{\dagger}$Center for Educational Assistance\\
  Shibaura Institute of Technology\\
  Saitama, 337--8570\\
  Japan\\
  }
\date{}
\maketitle
%\thispagestyle{empty}
%
%
%  gaiyou
%
%
\begin{abstract}
  In this paper the explicit flow representation to the Bose--Hubbard 
  Hamiltonian is given in the general case. This representation may be 
  useful in creating cat states for the system of atoms trapped in the 
  optical ring.
\end{abstract}
%

%\newpage

%
%
%     Honbun
%
%

To make macroscopic superpositions (a kind of cat states) of superfluid flows 
in a Bose--Einstein condensate (BEC) is really an interesting object in 
Quantum Mechanics. As a general introduction to BECs see for example 
\cite{PSm} and \cite{PSt}. When studying such states theoretically we usually 
use the Bose--Hubbard model, which is convenient enough for our purpose.

We are especially interested in BECs trapped in an optical lattice in a ring 
geometry. On this subject there are many works, see \cite{five}, or 
\cite{HBD-1}, \cite{DH}, \cite{HBD-2} and their references.

In studying the model we often use a Fourier transform, which is a fundamental 
tool. In \cite{HBD-1}, \cite{DH}, \cite{HBD-2} the flow representation (a 
discrete Fourier transform) of the Bose--Hubbard Hamiltonian with three BECs 
is given to study a cat state. However, the system of three BECs is too small 
for our purpose.

We are studying a quantum computation based on Cavity QED (see \cite{FHKW1} 
and \cite{FHKW2}) and want to construct a hybrid system consisting of 
our method and some ideas (devices) coming from BECs, so it is important 
to know the flow representation of the Hamiltonian {\bf in the general case}.

In this report we give the explicit flow representation to the Bose--Hubbard 
Hamiltonian in the general case. 
Our work may give a speed-up in studying the subject.

\vspace{5mm}
We treat in the paper the system of $n$ harmonic oscillators 
$\{(a_{1},a_{1}^{\dagger}),(a_{2},a_{2}^{\dagger}),\cdots,
(a_{n},a_{n}^{\dagger})\}$
\[
[a_{i},a_{j}^{\dagger}]=\delta_{ij},\quad 
[a_{i},a_{j}]=0,\quad 
[a_{i}^{\dagger},a_{j}^{\dagger}]=0
\]
trapped in an optical lattice in a ring geometry, see the following figure.
\vspace{5mm}
\begin{center}
\input{ring-geometry.fig}
\end{center}
\begin{center}
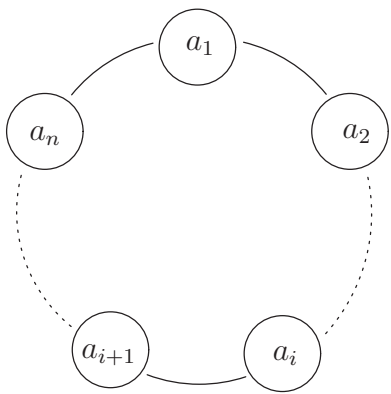
FIG : $n$ atoms trapped in an optical lattice in a ring geometry
\end{center}

The Bose--Hubbard Hamiltonian in our context is
\begin{equation}
\label{eq:bose-hubbard I}
\mbox{H}=-J\left(\sum_{i=1}^{n}a_{i+1}^{\dagger}a_{i}+h.c.\right)
+\frac{U}{2}\sum_{i=1}^{n}{a_{i}^{\dagger}}^{2}a_{i}^{2}
\end{equation}
where $a_{n+1}=a_{1}$ (mod $n$) and $J$, $U$ are some coupling constants 
consisting of the tunnelling strength, the interaction one respectively. 
See \cite{five} or \cite{HBD-2} and its references.

Let us rewrite the Hamiltonian by use of the matrix (vector) expression. 
We set 
\[
{\bf a}=(a_{1},a_{2},\cdots,a_{n-1},a_{n})^{t}
\]
where $t$ means the transpose and 
\begin{equation}
\label{eq:generalized Pauli}
\Sigma_{1}=
\left(
\begin{array}{cccccc}
0 &   &        &        &   & 1  \\
1 & 0 &        &        &   &    \\
  & 1 & 0      &        &   &    \\
  &   & \ddots & \ddots &   &    \\
  &   &        & 1      & 0 &    \\
  &   &        &        & 1 & 0
\end{array}
\right), \qquad
\Sigma_{3}=
\left(
\begin{array}{cccccc}
1 &        &            &        &                &                \\
  & \sigma &            &        &                &                \\
  &        & {\sigma}^2 &        &                &                \\
  &        &            & \ddots &                &                \\
  &        &            &        & {\sigma}^{n-2} &                \\
  &        &            &        &                & {\sigma}^{n-1}
\end{array}
\right)
\end{equation}
where $\sigma=\exp(2\pi\sqrt{-1}/n)$ which satisfies the relations
\[
\sigma^{n}=1,\quad \bar{\sigma}=\sigma^{n-1},\quad 
1+\sigma+\cdots +\sigma^{n-1}=0.
\]

The generators $\{\Sigma_{1},\Sigma_{3}\}$ are called the generalized Pauli 
matrices, which play an crucial role in qudit theory, see for example 
\cite{KF1}, \cite{FFK}, \cite{KF2}. 
Then we have easily
\begin{equation}
\label{eq:bose-hubbard II}
\mbox{H}=-J\left({\bf a}^{\dagger}\Sigma_{1}{\bf a}+h.c.\right)
+\frac{U}{2}\sum_{i=1}^{n}{a_{i}^{\dagger}}^{2}a_{i}^{2}.
\end{equation}

What we want to do is to make the first term in (\ref{eq:bose-hubbard II}) 
diagonal. For that let us remind the well--known decomposition
\begin{equation}
\label{eq:diagonal-form}
\Sigma_{1}=W\Sigma_{3}W^{\dagger}\ ;\quad 
\Sigma_{1}^{\dagger}=W^{\dagger}\Sigma_{3}W\quad 
(W^{-1}=W^{\dagger})
\end{equation}
with the generalized Walsh--Hadamard matrix (transformation) $W$ defined by
\begin{eqnarray}
\label{eq:Walsh-Hadamard}
W&=&\frac{1}{\sqrt{n}}
\left(
\begin{array}{cccccc}
1 & 1            & 1               & \cdots & 1 & 1                    \\
1 & \sigma^{n-1} & \sigma^{2(n-1)} & \cdots & \sigma^{(n-2)(n-1)} & 
\sigma^{(n-1)^2} \\
1 & \sigma^{n-2} & \sigma^{2(n-2)} & \cdots & \sigma^{(n-2)^2} & 
\sigma^{(n-1)(n-2)} \\
\vdots & \vdots  & \vdots &        & \vdots & \vdots                   \\
1& \sigma^{2} & \sigma^{4}& \cdots & \sigma^{2(n-2)} & \sigma^{2(n-1)} \\
1& \sigma & \sigma^{2} & \cdots & \sigma^{n-2} & \sigma^{n-1}
\end{array}
\right),   \\
W^{\dagger}&=&\frac{1}{\sqrt{n}}
\left(
\begin{array}{cccccc}
1 & 1            & 1               & \cdots & 1 & 1                      \\
1 & \sigma & \sigma^{2} & \cdots & \sigma^{n-2} & \sigma^{n-1}           \\
1 & \sigma^{2} & \sigma^{4} & \cdots & \sigma^{2(n-2)} & \sigma^{2(n-1)} \\
\vdots & \vdots  & \vdots &        & \vdots & \vdots                     \\
1& \sigma^{n-2} & \sigma^{2(n-2)}& \cdots & \sigma^{(n-2)^2} & 
\sigma^{(n-2)(n-1)} \\
1& \sigma^{n-1} & \sigma^{2(n-1)} & \cdots & \sigma^{(n-2)(n-1)} & 
\sigma^{(n-1)^2}
\end{array}
\right).
\end{eqnarray}

An interesting property of $W$ used later is
\begin{equation}
\label{eq:square-W}
W^{2}={W^{\dagger}}^{2}=
\left(
\begin{array}{cccccc}
1 &   &   &         &   &    \\
  &   &   &         &   & 1  \\
  &   &   &         & 1 &    \\
  &   &   & \rddots &   &    \\
  &   & 1 &         &   &    \\
  & 1 &   &         &   &    
\end{array}
\right).
\end{equation}
It is worth noting that this property has been used in constructing the 
exchange gate (operator) in qudit theory (see \cite{KF1} where $K=W^{2}$).

A comment may be in order. In the case of $n=2$ we have
\[
W=\frac{1}{\sqrt{2}}
\left(
\begin{array}{cc}
1 & 1  \\
1 & -1
\end{array}
\right), \quad
W^{2}=
\left(
\begin{array}{cc}
1 &   \\
  & 1
\end{array}
\right)
={\bf 1}_{2}.
\]
Surely, there is a (big) difference between $n=2$ and $n\geq 3$.

Now we set 
\begin{equation}
\label{eq:DFT}
{\bf \alpha}=W^{\dagger}{\bf a}\ \Longleftrightarrow\ 
{\bf a}=W{\bf \alpha},
\quad 
{\bf \alpha}=(\alpha_{1},\alpha_{2},\cdots,\alpha_{n-1},\alpha_{n})^{t},
\end{equation}
or more explicitly
\[
\alpha_{1}=\frac{1}{\sqrt{n}}\sum_{i=1}^{n}a_{i},\quad
\alpha_{2}=\frac{1}{\sqrt{n}}\sum_{i=1}^{n}\sigma^{i-1}a_{i},\quad
\cdots,\quad
\alpha_{n}=\frac{1}{\sqrt{n}}\sum_{i=1}^{n}\sigma^{(i-1)(n-1)}a_{i}.
\]
This change of ``variables" is a discrete Fourier transform in quantum 
computation and is called the flow representation \footnote{we don't know 
whether this terminology is universal or not}, 
\ see \cite{HBD-1}, \cite{DH}, \cite{HBD-2}.

Under this transform the first term of (\ref{eq:bose-hubbard II}) 
becomes
\begin{equation}
-2J\sum_{i=1}^{n}\cos\left(\frac{2\pi(i-1)}{n}\right)
\alpha_{i}^{\dagger}\alpha_{i},
\end{equation}
which is just diagonal as required. 

The real problem is to determine the second term of 
(\ref{eq:bose-hubbard II}) in terms of (\ref{eq:DFT}). 
In fact, it is not so easy. The aim of this paper is to give the explicit 
form to the term and we obtain the following
\vspace{3mm}
\par \noindent 
{\bf Fundamental Formula}
\begin{equation}
\sum_{i=1}^{n}{a_{i}^{\dagger}}^{2}a_{i}^{2}
=
\frac{1}{n}\sum_{j=0}^{n-1}
\left({\bf \alpha}^{t}W^{2}\Sigma_{1}^{j}{\bf \alpha}\right)^{\dagger}
\left({\bf \alpha}^{t}W^{2}\Sigma_{1}^{j}{\bf \alpha}\right).
\end{equation}

\vspace{3mm}
The proof will be given in \cite{FS}. It is interesting to note that $W^{2}$ 
in (\ref{eq:square-W}) is used in an essential manner. 

\par \vspace{3mm} \noindent
Therefore the Bose--Hubbard Hamiltonian (\ref{eq:bose-hubbard I}) 
becomes 
\begin{equation}
\label{eq:flow-representation}
\mbox{H}=-2J\sum_{i=1}^{n}\cos\left(\frac{2\pi(i-1)}{n}\right)
\alpha_{i}^{\dagger}\alpha_{i}+
\frac{U}{2n}\sum_{j=0}^{n-1}
\left({\bf \alpha}^{t}W^{2}\Sigma_{1}^{j}{\bf \alpha}\right)^{\dagger}
\left({\bf \alpha}^{t}W^{2}\Sigma_{1}^{j}{\bf \alpha}\right)
\end{equation}
in the {\bf flow representation}.

Let us write down (\ref{eq:flow-representation}) explicitly 
for some special cases ($n=3$, $4$ and $5$).

\par \noindent
{\bf n=3\ :}\ (\cite{HBD-1}, \cite{DH}, \cite{HBD-2})
\begin{eqnarray}
\label{eq:n=3}
\mbox{H}&=&-J\left(2\alpha_{1}^{\dagger}\alpha_{1}-
\alpha_{2}^{\dagger}\alpha_{2}-\alpha_{3}^{\dagger}\alpha_{3}\right)+
\frac{U}{6}
\left\{
\left((\alpha_{1}^{\dagger})^{2}+2\alpha_{2}^{\dagger}\alpha_{3}^{\dagger}
\right)\left(\alpha_{1}^{2}+2\alpha_{2}\alpha_{3}\right)
\right. \nonumber \\
&&
\left.
+
\left((\alpha_{2}^{\dagger})^{2}+2\alpha_{1}^{\dagger}\alpha_{3}^{\dagger}
\right)\left(\alpha_{2}^{2}+2\alpha_{1}\alpha_{3}\right)
+
\left((\alpha_{3}^{\dagger})^{2}+2\alpha_{1}^{\dagger}\alpha_{2}^{\dagger}
\right)\left(\alpha_{3}^{2}+2\alpha_{1}\alpha_{2}\right)
\right\}.
\end{eqnarray}

\par \noindent
{\bf n=4\ :}
\begin{eqnarray}
\label{eq:n=4}
\mbox{H}&=&-2J\left(\alpha_{1}^{\dagger}\alpha_{1}-
\alpha_{3}^{\dagger}\alpha_{3}\right)+
\frac{U}{8}
\left\{
\left((\alpha_{1}^{\dagger})^{2}+2\alpha_{2}^{\dagger}\alpha_{4}^{\dagger}+
(\alpha_{3}^{\dagger})^{2}\right)
\left(\alpha_{1}^{2}+2\alpha_{2}\alpha_{4}+\alpha_{3}^{2}\right)
\right. \nonumber \\
&&
\left.
+
\left((\alpha_{2}^{\dagger})^{2}+2\alpha_{1}^{\dagger}\alpha_{3}^{\dagger}+
(\alpha_{4}^{\dagger})^{2}\right)
\left(\alpha_{2}^{2}+2\alpha_{1}\alpha_{3}+\alpha_{4}^{2}\right)
+
4\left(\alpha_{1}^{\dagger}\alpha_{2}^{\dagger}+
\alpha_{3}^{\dagger}\alpha_{4}^{\dagger}\right)
\left(\alpha_{1}\alpha_{2}+\alpha_{3}\alpha_{4}\right)
\right. \nonumber \\
&&
\left.
+
4\left(\alpha_{1}^{\dagger}\alpha_{4}^{\dagger}+
\alpha_{2}^{\dagger}\alpha_{3}^{\dagger}\right)
\left(\alpha_{1}\alpha_{4}+\alpha_{2}\alpha_{3}\right)
\right\}.
\end{eqnarray}

\par \noindent
{\bf n=5\ :}
\begin{eqnarray}
\label{eq:n=5}
\mbox{H}&=&-\frac{J}{2}\left\{
4\alpha_{1}^{\dagger}\alpha_{1}+
(\sqrt{5}-1)\left(\alpha_{2}^{\dagger}\alpha_{2}+
\alpha_{5}^{\dagger}\alpha_{5}\right)-
(\sqrt{5}+1)\left(\alpha_{3}^{\dagger}\alpha_{3}+
\alpha_{4}^{\dagger}\alpha_{4}\right)
\right\}  \nonumber \\
&&+\frac{U}{10}
\left\{
\left((\alpha_{1}^{\dagger})^{2}+2\alpha_{2}^{\dagger}\alpha_{5}^{\dagger}+
2\alpha_{3}^{\dagger}\alpha_{4}^{\dagger}\right)
\left(\alpha_{1}^{2}+2\alpha_{2}\alpha_{5}+2\alpha_{3}\alpha_{4}\right)+
\right. \nonumber \\
&&\qquad\quad
\left((\alpha_{3}^{\dagger})^{2}+2\alpha_{1}^{\dagger}\alpha_{5}^{\dagger}+
2\alpha_{2}^{\dagger}\alpha_{4}^{\dagger}\right)
\left(\alpha_{3}^{2}+2\alpha_{1}\alpha_{5}+2\alpha_{2}\alpha_{4}\right)+
\nonumber \\
&&\qquad\quad 
\left((\alpha_{5}^{\dagger})^{2}+2\alpha_{1}^{\dagger}\alpha_{4}^{\dagger}+
2\alpha_{2}^{\dagger}\alpha_{3}^{\dagger}\right)
\left(\alpha_{5}^{2}+2\alpha_{1}\alpha_{4}+2\alpha_{2}\alpha_{3}\right)+
\nonumber \\
&&\qquad\quad 
\left((\alpha_{2}^{\dagger})^{2}+2\alpha_{1}^{\dagger}\alpha_{3}^{\dagger}+
2\alpha_{4}^{\dagger}\alpha_{5}^{\dagger}\right)
\left(\alpha_{2}^{2}+2\alpha_{1}\alpha_{3}+2\alpha_{4}\alpha_{5}\right)+
\nonumber \\
&&\qquad\quad 
\left.
\left((\alpha_{4}^{\dagger})^{2}+2\alpha_{1}^{\dagger}\alpha_{2}^{\dagger}+
2\alpha_{3}^{\dagger}\alpha_{5}^{\dagger}\right)
\left(\alpha_{4}^{2}+2\alpha_{1}\alpha_{2}+2\alpha_{3}\alpha_{5}\right)
\right\}.
\end{eqnarray}

\vspace{5mm}
In this paper we treated the Bose--Hubbard Hamiltonian consisting of atoms 
trapped in the optical ring and gave the explicit flow representation 
(discrete Fourier transform) to it. 
This representation may be convenient in creating cat states for the system, 
which will be reported in another paper \cite{FS}.

We conclude the paper by stating our motivation once more. We are studying 
a quantum computation (computer) based on Cavity QED, so to construct a more 
realistic model of (robust) quantum computer we have to study a hybrid system 
consisting of our method and some ideas (devices) coming from BECs. 
This is our future task.

%%%%%%%%%%%%%
%References%
%%%%%%%%%%%%%

\end{document}